\documentclass[prl,graphicx,twocolumn,showpacs,superscriptaddress]{revtex4-1}

\usepackage{amsmath,amssymb,amscd}
\usepackage{enumerate}
\usepackage{graphicx}
\usepackage{mathbbol}
\usepackage{amsfonts}
\usepackage{natbib}

\usepackage{bm}

\usepackage{color}

\draft 

\makeatletter
\newcommand{\raisemath}[1]{\mathpalette{\raisem@th{#1}}}
\newcommand{\raisem@th}[3]{\raisebox{#1}{$#2#3$}}
\makeatother

\begin{document}

\title{Exact Statistics of the Gap and Time Interval Between the First Two Maxima of Random Walks}

\author{Satya N. Majumdar}
\email[]{majumdar@lptms.u-psud.fr}
\affiliation{Univ. Paris-Sud, CNRS, LPTMS, 91405 Orsay Cedex, France}

\author{Philippe Mounaix}
\email[]{philippe.mounaix@cpht.polytechnique.fr}
\affiliation{Centre de Physique Th\'eorique, UMR 7644 CNRS, Ecole Polytechnique, 91128 Palaiseau Cedex, France}

\author{Gr\'egory Schehr}
\email[]{gregory.schehr@lptms.u-psud.fr}
\affiliation{Univ. Paris-Sud, CNRS, LPTMS, 91405 Orsay Cedex, France}

\date{\today}

\begin{abstract}
We investigate the statistics of the gap, $G_n$, between the two rightmost positions of a Markovian one-dimensional random walker (RW) after $n$ time steps and of the duration, $L_n$, which separates the occurrence of these two extremal positions. The distribution of the jumps $\eta_i$'s of the RW, $f(\eta)$, is symmetric and its Fourier transform has the small $k$ behavior $1-\hat{f}(k)\sim\vert k\vert^\mu$ with $0 < \mu \leq 2$. We compute the joint probability density function (pdf) $P_n(g,l)$ of $G_n$ and $L_n$ and show that, when $n \to \infty$, it approaches a limiting pdf $p(g,l)$.  
The corresponding marginal pdf of the gap, $p_{\rm gap}(g)$, is found to behave like $p_{\rm gap}(g) \sim g^{-1 - \mu}$ for $g \gg 1$ and $0<\mu < 2$. We show that the limiting marginal distribution of $L_n$, $p_{\rm time}(l)$, has an algebraic tail $p_{\rm time}(l) \sim l^{-\gamma(\mu)}$ for $l \gg 1$ with $\gamma(1<\mu \leq 2) = 1 + 1/\mu$, and $\gamma(0<\mu<1) = 2$. For $l, g \gg 1$ with fixed $l g^{-\mu}$, $p(g,l)$ takes the scaling form $p(g,l) \sim g^{-1-2\mu} \tilde p_\mu(l g^{-\mu})$ where $\tilde p_\mu(y)$ is a ($\mu$-dependent) scaling function. We also present numerical simulations which verify our analytic results.
%
\end{abstract}


\maketitle 

Extreme and order statistics are currently the subject of numerous studies in various areas of sciences. In many circumstances, statistical systems are not governed by typical or average events but instead by anomalously rare and intense ones \cite{Gum58}. Illustrative examples of such situations include for instance natural disasters \cite{KPN02} or financial crisis \cite{EKM97} where extreme events, like earthquakes, tsunamis or financial crashes may have drastic consequences. It is also now well established that extreme value questions play an important role in the statistical mechanics of disordered systems \cite{BM97,DM01,LDM03}. 

When studying extreme statistics, one is usually interested in studying the maximum $X_{\rm max}$ among a collection of $N+1$ random variables $X_0, X_1, \cdots, X_N$. However, the knowledge of the statistics of this single global variable, though important, does not always provide enough valuable information. In particular, a crucial question concerns the crowding of events near the maximum. For instance, a rare event like an earthquake is usually not isolated but is followed (or preceded) by smaller ones, called aftershocks (respectively foreshocks) \cite{Omo1894, Uts61}. Foreshocks and aftershocks are also known to occur before and after a financial market shock \cite{LM03, PWHS10}. This is a natural question in statistical physics too, when one is interested not only in the ground state properties of a disordered system but also in the finite low temperature physics, which involves the low lying energy states, close to the ground state \cite{MLD04}. This has led to the study of the density of near extreme events, both in statistics \cite{PL98} and in physics \cite{SM07}, which essentially counts the number of events $X_i$'s which are at given distance from $X_{\rm max}$~\cite{SM07}. Crowding of near extreme events has also been studied in the context of sporting events, like marathon packs \cite{SMR2008}. 

Another natural way to characterize this crowding phenomenon is to study the order statistics of the $X_i$'s, that is the statistics of $X_{\rm max} = M_{1,n} > ... > M_{k,n} > ... > M_{n+1,n}$ where $M_{k,n}$ denotes the $k^{\rm th}$ maximum of the set $\{X_0, X_1,\dots, X_n\}$. One natural question for disordered systems, if one interprets $-M_{k,n}$ as the $k^{\rm th}$ energy level of the system, is the distribution of the first gap $G_n = M_{1,n} - M_{2,n}$ as it controls, to a large extent, the low temperature properties of the system. The gap $G_n$ is also an important quantity for applications in seismology as it can model the difference in magnitude between the mainshock and its largest aftershock or foreshock \cite{Bat65,Ver69,CLMR03}. 
%
%
%
%
%
\begin{center}
\begin{figure}[hht]
\includegraphics[width=\linewidth]{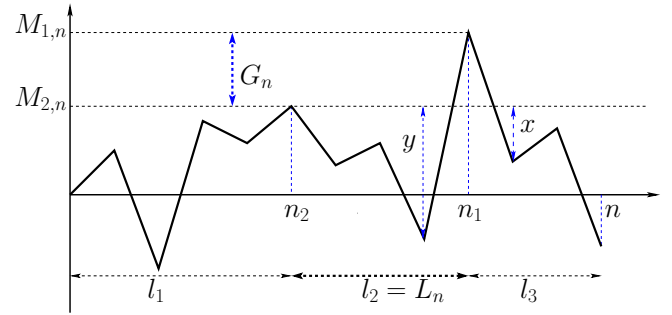}
\caption{Realization of a random walk (\ref{eq:Markov}) of $n$ steps. Here we focus on the joint pdf $P_n(g,l)$ of $G_n$ and $L_n$ in the limit of large $n$.}\label{Fig_markov}
\end{figure}
\end{center}
Aside from the gap, another key random variable is the time $L_n$ elapsed between these two extreme events (see Fig. \ref{Fig_markov}), which is particularly important for the statistics of earthquakes or financial crashes. 

While the study of the fluctuations of $G_n$ and $L_n$ is very well understood in the case of independent and identically distributed (i.i.d.) random variables \cite{order_book}, this question is highly non trivial for strongly correlated variables $X_i$'s. Yet it is known, for instance, that aftershocks exhibit long range correlations \cite{Omo1894,Uts61,YBZ09} (both spatial and temporal), in which case a model of independent variables cannot predict anything sensible about the distribution of the time separating the mainshock from the largest aftershock. The importance of order statistics for strongly correlated variables came up recently in several other physical contexts, notably in the study of the branching Brownian motion (BBM)~\cite{derrida_bbm} and also for $1/f^\alpha$ signals~\cite{racz_order}. In the former case, $G_n$ represents the gap between the two rightmost particles. The pdf of $G_n$ was studied in~\cite{derrida_bbm}, though an exact analytical expression remains a hard task. Therefore any exact results for the statistics of the gap and the time between the two first maxima for a set of strongly correlated variables would be highly desirable.

In this Letter, we obtain exact results and find a very rich behavior for the statistics of $G_n$ and $L_n$ in the case where the $X_i$'s correspond to the positions of a random walker (RW) at discrete times $i$'s. Such RW is certainly the simplest, yet non trivial, set of strongly correlated random variables for which interesting problems of records \cite{MZ08} and order statistics \cite{SM12} can be solved exactly. Hence, while RW might not be a realistic model for earthquakes, this is a useful laboratory where the effects of correlations on the statistics of $G_n$ and $L_n$ can be studied in detail.

In our model, the RW starts at $X_0=0$ at time $0$ and evolves via 
\begin{eqnarray}
X_i= X_{i-1} +\eta_i \;, \label{eq:Markov}
\end{eqnarray}
where the $\eta_i$'s are i.i.d. random jumps each drawn from a symmetric distribution $f(\eta)$. Its
Fourier transform, $\hat f(k) = \int_{-\infty}^\infty  e^{ik\eta} f(\eta) d\eta$, has the small $k$ behavior
\begin{eqnarray}\label{eq:small_k}
\hat f(k) = 1 - |a k|^\mu + o(|k|^\mu) \;, \; 0 < \mu \leq 2 \;,
\end{eqnarray}
where $a$ is the characteristic length scale of the jumps. In particular, for $0<\mu < 2$, one has $f(\eta) \sim B/|\eta|^{\mu +1}$, for large $|\eta|$, with $B = a^\mu \sin{(\mu \pi/2) \Gamma[\mu +1]/\pi}$. Let $M_{1,n}$ and $M_{2,n}$ respectively denote the first and second maxima of the random walk~(\ref{eq:Markov}) after $n$ steps and write $n_1$ and $n_2$ the times at which they are reached: ${X}_{n_1} = M_{1,n}$ and $X_{n_2} = M_{2,n}$. The purpose of this work is to study the joint pdf $P_n(g,l)$ of the gap $G_n = M_{1,n} - M_{2,n}$ and the time $L_n = n_1 - n_2$ between the occurrence of these first two maxima (Fig. \ref{Fig_markov}). 


It is useful to summarize our main results. We first show that $P_n(g,l)$ has a well defined limiting  pdf $p(g,l)$ as $n \to \infty$. By integration over $l$ one obtains an exact expression for the marginal distribution $p_{\rm gap}(g)$ (\ref{eq:I1_I2}, \ref{eq:dist_gap}) the full form of which depends on $f(\eta)$.  For $0 <\mu < 2$, we show that $p_{\rm gap}(g)$ has an algebraic tail
\begin{eqnarray}\label{eq:result_p_of_g}
p_{\rm gap}(g) \sim {B_\mu} \; {g^{-1-\mu}}  \;, \; g \to \infty \;,
\end{eqnarray}
where $B_\mu$ is a computable constant. For $\mu = 2$, the full distribution can be computed in some specific cases only, and the tail itself remains non-universal and sensitive to $f(\eta)$. By integration over $g$, one finds that the marginal distribution $p_{\rm time}(l)$ displays an algebraic tail whose exponent depends only on the L\'evy index $\mu$ as follows:  
 \begin{equation}\label{eq:result_p_of_l}
p_{\rm time}(l) \sim 
\begin{cases}
{A_{\rm I}}\,{l^{-1-1/\mu}} & \;, \; 1 < \mu \leq 2 \\
{A_{\rm II} \, (\log l)}\,{l^{-2}} & \;, \, \mu = 1 \\
{A_{\rm III}}\,{l^{-2}} & \;, \; 0 <\mu\ne\frac{1}{k} < 1 \;,
\end{cases}
\end{equation}
for $l\to\infty$, where the amplitudes $A_{\rm I}, A_{\rm II}$ and $A_{\rm III}$ are non universal and $k \in {\mathbb N}$ \cite{foot1}. The third line of\ (\ref{eq:result_p_of_l}) reveals an unexpected freezing phenomenon of the exponent characterizing the algebraic tail of $p_{\rm time}(l)$ as $\mu$ decreases past the value $\mu_c = 1$. Interestingly, we see that the first moment of $p_{\rm time}(l)$ is not defined. This means that although the typical size of $L_n$ is ${\cal O}(1)$, its average diverges with $n$. From (\ref{eq:result_p_of_l}) one can estimate that $\langle L_n \rangle \sim n^{1-1/\mu}$ for $1< \mu \le 2$, while $\langle L_n \rangle \sim \log{n}$ for $0 < \mu < 1$. Finally, in the scaling regime $g,\ l \gg 1$ with fixed $l g^{-\mu}$ and for $0 < \mu < 2$, we find the following scaling form
\begin{eqnarray}\label{eq:scaling_jpdf}
p(g,l) \sim g^{-1-2\mu} \tilde p_\mu(l g^{-\mu}) \;,
\end{eqnarray}
where the $\mu$-dependent scaling function $\tilde p_\mu(y)$ is integrable over $\lbrack 0,+\infty)$ with the algebraic tail
\begin{eqnarray}\label{eq:asympt_scaling_jpdf}
\tilde p_\mu(y) \sim y^{-1-1/\mu} \;, \; y \to +\infty \;.
\end{eqnarray}

The starting point of our analysis is an exact formula for the joint pdf of the random variables $n_1$, $n_2$, and $G_n = M_{1,n}-M_{2,n}$ for a random walk of $n$ steps, in terms of the following two central objects. The first one is the survival probability $q_n(x)$ for a random walker, starting at $x\ge 0$, to stay on the positive axis up to step $n$. Note that by $q_n(x) = {\Pr}(M_{1,n} \leq x)$, it is also the cumulative distribution of $M_{1,n}$ \cite{CM05}. A complete characterization of $q_n(x)$ is given by the Laplace transform (LT) with respect to (wrt) $x$ of its generating function (GF) wrt $n$~\cite{Pol75,Spi56} (see also \cite{Maj10}),
\begin{eqnarray}\label{eq:Poll-Spitz}
\int_0^\infty dx \sum_{m=0}^\infty q_m(x) s^m e^{-\lambda x} = \frac{1}{\lambda \sqrt{1-s}} \varphi(s,\lambda) \;,
\end{eqnarray}  
where the function $\varphi(s,\lambda)$ is given by
\begin{eqnarray}\label{eq:phi}
\varphi(s,\lambda) = \exp{\left(-\frac{\lambda}{\pi} \int_0^\infty \frac{\ln{\lbrack 1-s \hat f(k) \rbrack}}{k^2 + \lambda^2} \, dk\right)} \;.
\end{eqnarray}
The second object is the probability $p_n(x) dx$ for a random walker, starting at $x=0$, and conditioned to stay positive, to arrive at step $n$ in the interval $[x,x+dx]$. The counterpart of~(\ref{eq:Poll-Spitz}) for $p_n(x)$ reads~\cite{Iva94} (see also \cite{MCZ06})
\begin{eqnarray}\label{eq:Ivanov}
\int_0^\infty dx \sum_{m=0}^\infty p_m(x) s^m e^{-\lambda x} = \varphi(s,\lambda) \;.
\end{eqnarray}
To compute $P(g, n_1, n_2 | n)$, we exploit the renewal property of the random walk and divide the interval $\lbrack 0,n\rbrack$
into three independent parts (see Fig. \ref{Fig_markov}) of duration $l_1 = n_2$, $l_2 = n_1 - n_2$ and $l_3 = n-n_1$, (here we suppose without loss of generality that $n_1 > n_2$). One has,
\begin{eqnarray}\label{eq:renewal}
P(g, n_1, n_2 | n) = w_1(l_1) w_2(l) w_3(l_3) \delta_{l_1+l+l_3,n} \;,
\end{eqnarray}
where $w_k(l_k)$, with $k = 1,2,3$ and $l_2=l$, denotes the weight of the paths in each of the three subintervals (see Fig. \ref{Fig_markov}) and $\delta_{i,j}$ is the Kronecker delta function. The weight $w_1(l_1)$ is simply given by the survival probability
\begin{eqnarray}\label{eq:w1}
w_1(l_1) = q_{\raisemath{-2pt}{l_1}}(0) \;.
\end{eqnarray} 
This can be readily seen by reversing the direction of both space and time axis and taking as a new origin the point of coordinates $(l_1,M_{2,n})$ (notice that $P(g,n_1,n_2|n)$ is obtained by integrating over the value of $M_{2,n}$ keeping the value of the gap $g$ fixed). To compute $w_2(l)$ we isolate the last step, of amplitude $g + y$, before the maximum $M_{1,n}$ is reached, from the first $n_2-n_1-1$ steps on this interval (Fig. \ref{Fig_markov}). The weight associated to these $n_2-n_1-1$ steps is given by $p_{n_2-n_1-1}(y)$ and hence
\begin{eqnarray}\label{eq:w2}
w_2(l) = \int_0^\infty p_{l-1}(y) f(g+y) \, dy \;.
\end{eqnarray} 
Similarly, to compute the weight $w_3(l_3)$ we isolate the first step, of amplitude $g+x$,  after the maximum $M_{1,n}$ is reached from the last $n-n_1-1$ steps. The weight associated to these last steps is simply given by $q_{n-n_1-1}(x)$ and one has
\begin{eqnarray}\label{eq:w3}
w_3(l_3) = \int_0^\infty q_{\raisemath{-2pt}{l_3-1}}(x) f(g+x) \, dx \;.
\end{eqnarray}

The joint pdf $P_n(g,l)$ is obtained formally from~(\ref{eq:renewal}) as $P_n(g,l)=\sum_{l_1,l_3=0}^{+\infty}P(g, n_1, n_2 | n)$, and using (\ref{eq:w1})-(\ref{eq:w3}) together with (\ref{eq:Poll-Spitz}) and (\ref{eq:Ivanov}), one obtains an explicit expression for the double GF of $P_n(g,l)$ wrt $n$ and $l$ [we recall that $P_n(g,l) = P_n(g,-l)$]. Namely,
\begin{eqnarray}\label{eq:joint_doubleLT}
\sum_{n= 0}^\infty s^n \sum_{l=0}^n t^l \, P_n(g,l)  &=& \frac{t s^2}{1-s} \int_0^\infty u(st,y) f(g+y) \, dy \nonumber \\
&\times& \int_0^\infty h(s,x) f(g+x) \, dx \;,
\end{eqnarray} 
where $u(t,y)$ and $h(s,x)$ are the inverse LTs of $\varphi(t,\lambda)$ and $\varphi(s,\lambda)/\lambda$, respectively:
\begin{equation}\label{eq:defu_and_h}
\begin{array}{l}
\int_0^\infty u(t,y) e^{-\lambda y} \, dy= \varphi(t,\lambda) \;, \\
\\
\int_0^\infty h(s,x) e^{-\lambda x} \, dx = \varphi(s,\lambda)/\lambda \;.
\end{array}
\end{equation}
From the $s \to 1$ limit of~(\ref{eq:joint_doubleLT}) one can extract the large $n$ limit of $P_n(g,l)$. It is readily seen that the leading divergence on the right hand side of (\ref{eq:joint_doubleLT}) is a simple pole at $s=1$: this implies that $P_n(g,l)$ converges to a limiting distribution $p(g,l)$ as $n \to \infty$ with GF wrt $l$
\begin{equation}\label{eq:gf_jpdf}
\tilde p(g,z) = \sum_{l=0}^\infty z^l p(g,l) = I_1(z,g) I_2(g) \\
\end{equation}
where
\begin{equation}\label{eq:I1_I2}
\begin{array}{l}
I_1(z,g) =  z \int_0^\infty u(z,y) f(g+y) \, dy \;, \\
\\
I_2(g) = \int_0^\infty h(1,x) f(g+x) \, dx \;.
\end{array}
\end{equation}
Expression (\ref{eq:gf_jpdf}), together with (\ref{eq:I1_I2}), is the central result of our study from which the various behaviors announced in the introduction can be derived. 

We first focus on the marginal distribution of the gap $p_{\rm gap}(g)$ which, for any jump distribution, is exactly given by
\begin{equation}\label{eq:dist_gap}
p_{\rm gap}(g) = 2 \tilde  p(g,1)= 2 I_1(1,g) I_2(g) \;,
\end{equation}
where the factor of $2$ comes from the configurations with $l\gtrless 0$. For $\mu=2$, there are some particular cases in which (\ref{eq:dist_gap}) can be computed explicitly. For instance, if $f(\eta) = (b/2) \exp{(-b |\eta|)}$ one finds $p_{\rm gap}(g) = 2 b \exp{(-2 b g)}$, and for $f(\eta) = (b^2/2) |\eta| \exp{(-b |\eta|)}$ one has $p_{\rm gap}(g) =c\lbrack(\sqrt{3}+ 2bg)^2-1\rbrack e^{-2 b g}$ with $c=2b/(1+\sqrt{3})^2$. Fig. {\ref{fig_gap} a) shows a numerical check of the latter exact result.
\begin{figure}
\includegraphics[width = \linewidth]{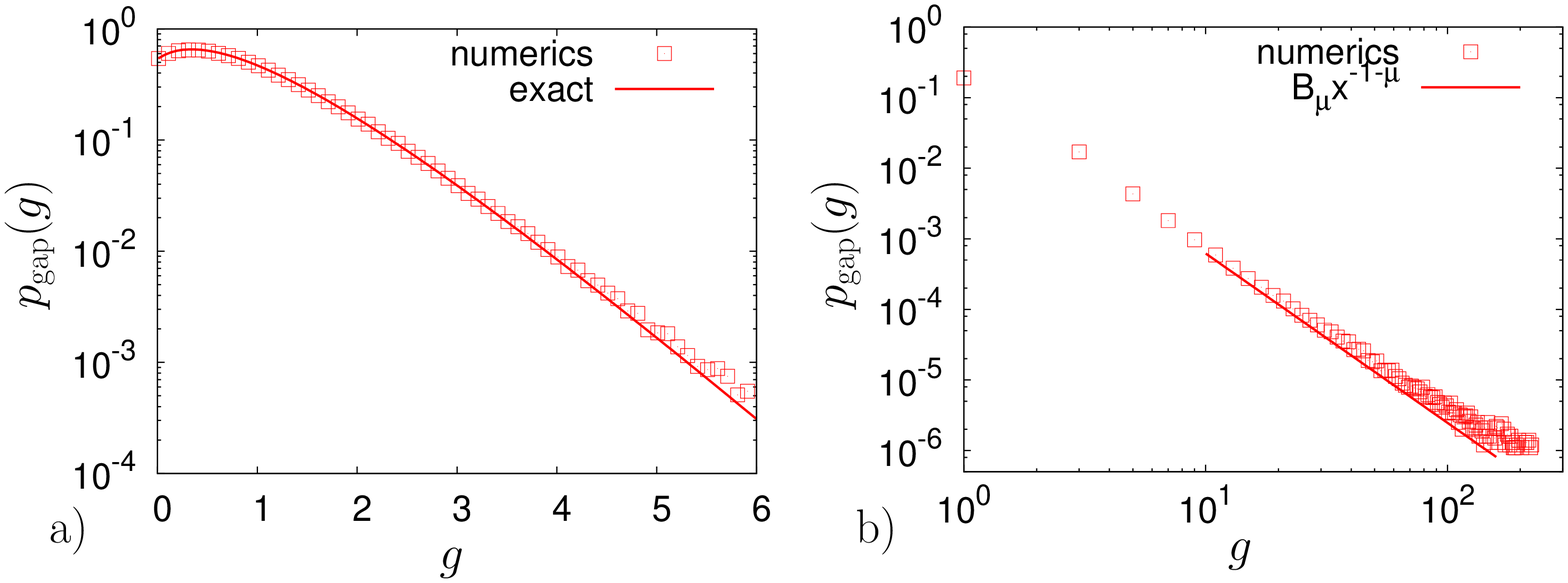}
\caption{{\bf a}: Plot of $p_{\rm gap}(g)$ for $f(\eta) = \frac{1}{2} |x| \exp{(- |x|)} $ and $n=10^4$. The dotted line corresponds to the exact result given in the text given below (\ref{eq:dist_gap}). {\bf b)}: Plot of $p_{\rm gap}(g)$ for $\mu = 1.4$ and $n=10^3$. The solid line is the exact asymptotic result.}\label{fig_gap}
\end{figure}
On these two examples we see that, for $\mu=2$, even the tail of $p_{\rm gap}(g)$ depends on the the details of the jump distribution. On the other hand, for
$0<\mu < 2$, the tail of $p_{\rm gap}(g)$ depends on the L\'evy index $\mu$ only. In the large $g$ limit, it turns out that the integrals over $x$ and $y$ in Eq. (\ref{eq:I1_I2}) are dominated by large values of $x,y \sim {\cal O}(g)$. Hence, to study the large $g$ limit, one needs the large argument behavior of $u(1,y)$ and $h(1,x)$ in Eq. (\ref{eq:defu_and_h}). These behaviors can in turn be obtained by analyzing the small $\lambda$ behavior of $\varphi(1,\lambda)$ in Eq. (\ref{eq:phi}). One finds that $u(1,y)\sim b_1 y^{\mu/2-1}$, with $b_1 = [a^{\mu/2} \Gamma(\mu/2)]^{-1}$, and $h(1,x) \sim c_1 x^{\mu/2}$ with $c_1 = [a^{\mu/2} \Gamma(1+\mu/2)]^{-1}$. Using these asymptotic behaviors, it can be shown from (\ref{eq:I1_I2}) and (\ref{eq:dist_gap}) that $p_{\rm gap}(g) \sim B_\mu \, g^{-1-\mu}$ as announced in Eq. (\ref{eq:result_p_of_g}) with the amplitude $B_\mu = a^\mu {\mu}/{[\Gamma(1-\mu/2)]^2}$. Fig. (\ref{fig_gap}) b) shows a numerical estimate of $p_{\rm gap}(g)$ for $\mu = 1.4$: the tail behavior is in good agreement with the analytical predictions (\ref{eq:result_p_of_g}). 

We now come to the marginal distribution $p_{\rm time}(l)$ of the time $l$ elapsed between the first two maxima. Its GF is readily obtained by integrating (\ref{eq:gf_jpdf}) over $g$. One finds,
\begin{eqnarray}\label{eq:gf_time}
\tilde p_{\rm time}(z) = \sum_{l=0}^\infty z^l p_{\rm time}(l) = \int_0^\infty dg I_1(z,g) I_2(g) \;.
\end{eqnarray} 
For an exponential jump distribution, $f(\eta) = (b/2) \exp{(-b|\eta|)}$, corresponding to $\mu=2$, $p_{\rm time}(l)$ can be computed exactly, $p_{\rm time}(l) =  \Gamma(l-\tfrac{1}{2})/[4 \sqrt{\pi} \Gamma(l+1)]$ and $p_{\rm time}(l) \propto l^{-3/2}$, for large $l$. For other jump distributions, an explicit computation of $p_{\rm time}(l)$ is generally impossible but its large $l$ behavior can be obtained from the $z \to 1$ limit of the GF (\ref{eq:gf_time}). This analysis is rather subtle and requires the analysis of $\varphi(z,\lambda)$ when $z \to 1$. In this limit, Eq. (\ref{eq:phi}) yields
\begin{equation}\label{eq:asympt_phi}
\varphi(z,\lambda) = \varphi(1,\lambda) + a_\mu \frac{(1-z)^{\frac{1}{\mu}}}{\lambda}
\varphi(1,\lambda) + o\left(\frac{(1-z)^{\frac{1}{\mu}}}{\lambda}\right) \;,
\end{equation}
with $a_\mu = (a \sin{(\pi/\mu)})^{-1}$. For $1< \mu \le 2$, the asymptotics (\ref{eq:asympt_phi}) can be directly used to compute the large $l$ behavior of $p_{\rm time}(l)$. One finds the first line of (\ref{eq:result_p_of_l}) with
\begin{equation}\label{eq:expr_aI}
A_{\rm I} = (a \pi)^{-1}\Gamma(1+1/\mu) \int_0^\infty \left[I_2(g)\right]^2 dg \;.
\end{equation}
A numerical check of these results for $1< \mu \leq 2$ is shown in Fig. \ref{fig_p_of_l}. For $\mu \le 1$ (and $\mu \neq 1/k$ with $k \in {\mathbb N}$ and $k>1$), things are more complicated since the $g$-integral defining $A_{\rm I}$ diverges (because $I_2(g) \sim g^{-\mu/2}$ for large $g$).
\begin{figure}[hh]
\includegraphics[width = \linewidth]{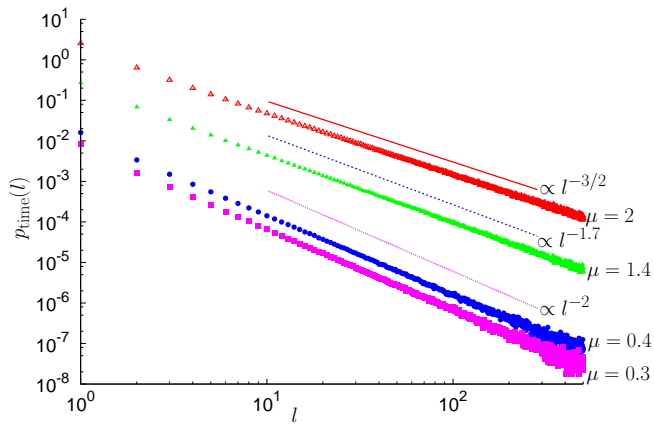}
\caption{{\bf Left}: Log-log plot of $p_{\rm time}(l)$ for different values of $\mu$ and $n=10^3$ (the case $\mu = 2$ corresponding to an exponential jump distribution). The straight lines correspond to an algebraic decay as predicted in Eq. (\ref{eq:result_p_of_l}) (note that $1+1/1.4 \approx 1.7$). These data have been vertically shifted for clarity.} 
\label{fig_p_of_l}
\end{figure}
In this regime, the large $l$ behavior of $p_{\rm time}(l)$ can be obtained by the following scaling argument. The expansion~(\ref{eq:asympt_phi}) holds for $(1-z)^{\frac{1}{\mu}}/ \lambda \ll 1$. This suggests that the $g$-integral in (\ref{eq:expr_aI}) should actually been cut-off around $l^{1/\mu}$ such that for $\mu < 1$, the second line of (\ref{eq:result_p_of_l}) is replaced with $p_{\rm time}(l) \sim A_{\rm III}l^{-2}$, hence the freezing of the tail of $p_{\rm time}(l)$ for $\mu < 1$, as discussed below Eq. (\ref{eq:result_p_of_l}). For the marginal case $\mu = 1$, one finds the logarithmic correction given in the second line of (\ref{eq:result_p_of_l}). Fig. \ref{fig_p_of_l} shows numerical results for $p_{\rm time}(l)$ with $\mu = 0.4$ and $\mu=0.3$ which both corroborate $p_{\rm time}(l) \propto l^{-2}$. 

From Eqs. (\ref{eq:result_p_of_g}) and (\ref{eq:result_p_of_l}) it is possible to get the scaling form of the joint pdf $p(g,l)$ for large $g$ and $l$ in (\ref{eq:scaling_jpdf}). According to standard scaling arguments, $p(g,l)$ is expected to depend on $l$ through the dimensionless combination $l g^{-\mu}$. Moreover, for large $g$ we showed that $p_{\rm gap}(g) \sim g^{-1 - \mu}$. From these two arguments, it is natural to expect that $p(g,l)$ has the scaling form (\ref{eq:scaling_jpdf}) where the function $\tilde{p}_\mu(y)$ is integrable over $[0,+\infty)$. Indeed, one can easily check that integrating (\ref{eq:scaling_jpdf}) over $l$ yields $p_{\rm gap}(g) \sim g^{-1 - \mu} \int_0^\infty \tilde{p}_\mu(y) dy $, in agreement with the large $g$ behavior $p_{\rm gap}(g) \sim g^{-1 - \mu}$ in Eq. (\ref{eq:result_p_of_g}), as it should. The large $y$ behavior of the scaling function $\tilde{p}_\mu(y)$ can be obtained from the large $l$ behavior of $p(g,l)$ at fixed $g$. Performing an analysis similar to the one leading to (\ref{eq:result_p_of_l}, \ref{eq:expr_aI}) one finds $p(g,l) \sim l^{-1 - 1/\mu} [I_2(g)]^2$ which, for large $g$, behaves like
\begin{eqnarray}\label{eq:jpdf_large_gl}
p(g,l)\sim g^{-\mu} l^{-1-1/\mu} \;, \;  l,g \to \infty \;.
\end{eqnarray}
It follows immediately that for large $y$, $\tilde p(y) \sim y^{-1 - \frac{1}{\mu}}$ as announced in Eq. (\ref{eq:asympt_scaling_jpdf}). This scaling form can be shown to be consistent with the large $l$ behavior of $p_{\rm time}(l)$ in Eq. (\ref{eq:result_p_of_l}). We have checked that Eq. (\ref{eq:asympt_scaling_jpdf}) is corroborated by numerical simulations with a good accuracy for different values of $\mu$. 
%
%

In this letter, we have investigated the statistical properties of the first gap, $G_n$, and the associated time interval between the two rightmost positions, $L_n$, for a one-dimensional random walk in the limit of infinitely many time steps. The scaling forms of the limiting pdf's $p(g,l)$, $p_{\rm gap}(g)$, and $p_{\rm time}(l)$ for large $g$ and $l$ have been obtained, see Eqs. (\ref{eq:result_p_of_g}) to (\ref{eq:asympt_scaling_jpdf}). Remarkable, unexpected, results are the freezing of the tail of $p_{\rm time}(l)$ to $p_{\rm time}(l)\sim l^{-2}$ for $0<\mu <1$, and the divergence of the average duration $\langle l \rangle=\int l\, p_{\rm time}(l)\, dl$ for any L\'evy index $0<\mu \le 2$. Moreover, while the first moment of the gap $\langle g \rangle = \int l\, p_{\rm time}(l)\, dg $ is finite for $1 < \mu \leq 2$, we found that it diverges for $0 < \mu < 1$. Such a divergence is at variance with the empirical B\aa th's law\ \cite{Bat65} for earthquakes that predicts a finite average
gap of magnitude between the mainshock and the next largest aftershock. While we focused here on the first gap, a recent study of the case $\mu =2$ \cite{SM12} showed that the $k^{\rm th}$ gap, with $k$ large (and $n \to \infty$), displays universal fluctuations with a power law tail. This naturally raises the question about the time between the $k^{\rm th}$ and $(k+1)^{\rm th}$ maximum in the limit of large $k$, which is a challenging problem. 

\acknowledgments{SNM and GS acknowledge support by ANR grant 2011-BS04-013-01 WALKMAT and in part by the Indo-French 
Centre for the Promotion of Advanced Research under Project 4604-3.}


\begin{thebibliography}{100}
\bibitem{foot1}
For $\mu = 1/k$, with $k \in \mathbb{N}$, there are some logarithmic corrections to the algebraic decay.
\end{thebibliography}
\end{document}